# A Mixed VR and Physical Framework to Evaluate Impacts of Virtual Legs and Elevated Narrow Working Space on Construction Workers' Gait Pattern

M. Habibnezhad[a], J. Puckett[a], M.S. Fardhosseini[b], L.A Pratama[b]

[a]Durham School of Architectural Engineering and Construction, University of Nebraska-Lincoln, Lincoln, NE, USA,68588
[b]Department of Construction Management, College of Built Environments, University of Washington, Seattle, WA, USA, 98195
E-mail: mahmoud@huskers.unl.edu , jay.puckett@unl.edu, sadrafh@uw.edu, pragung1@uw.edu

**Abstract –**
It is difficult to conduct training and evaluate workers' postural performance by using the actual job site environment due to safety concerns. Virtual reality (VR) provides an alternative to create immersive working environments without significant safety concerns. Working on elevated surfaces is a dangerous scenario, which may lead to gait and postural instability and, consequently, a serious fall. Previous studies showed that VR is a promising tool for measuring the impact of height on the postural sway. However, most of these studies used the treadmill as the walking locomotion apparatus in a virtual environment (VE). This paper was focused on natural walking locomotion to reduce the inherent postural perturbations of VR devices. To investigate the impact of virtual height on gait characteristics and keep the level of realism and feeling of presence at their highest, we enhanced the first-person-character model with 'virtual legs'. Afterward, we investigated its effect on the gait parameters of the participants with and without the presence of height. To that end, twelve healthy adults were asked to walk on a virtual loop path once at the ground level and once at the 17th floor of an unfinished structure. By quantitatively comparing the participants' gait pattern results, we observed a decrease in the stride length and increase in the gait duration of the participants exposed to height. At the ground level the use of enhanced model reduced participants' average stride length and height. In other words, in the presence of VR legs, the level of realism significantly increases and thus results in a better virtual gait simulation. The results of this study help us understand users' behaviors when they were exposed to elevated surfaces and establish a firm ground for gait stability analysis for the future height-related VR studies. We expect this developed VR platform can generate reliable results of VR application in more construction safety studies.

## 1 Introduction

According to the annual Bureau of Label Statistics (BLS) report, the construction industry contributed more than 10 percent of all fatal occupational injuries in 2017 [1]. Falling from elevated surfaces accounted for a significant portion of these incidents. According to Hsiao and Simonov [2], the primary stated reasons for falling were slips, trips, and loss of balance based on the worker compensation descriptions and fatality investigation reports. To address the disproportional rate of fall injuries and fatalities, researchers have been focusing on the influential factors affecting postural sway and gait stability. Many of these experiments, especially those related to fear of height and flight, were expensive and dangerous [3]. Therefore, recent studies have been investigating alternative methods to conduct postural stability and fall-related experiments. They found virtual reality (VR) as an alternative solution which "offers the opportunity to bring the complexity of the physical world into the controlled environment of the laboratory [4]."

VR is "a simulation in which computer graphics are used to create a realistic-looking world" [5]. Powerful real-time interaction capability, immersion feeling, and a much more intuitive link between the computer and the user turn this technology into one of the most rapidly growing technological fields. Specifically, there are extensive areas of VR applications ranging from manufacturing design and operation management— things like prototyping, product design, planning, and simulation [6]—to training and rehabilitation [4]. VR provides the users with an immersive environment, a wide range of view, and haptic feedback, delivering a firm platform for experiment designs in a much cheaper and safer way. Despite the inherent instability induction of VR headsets [7], this new technology has been used in numerous postural stability and fall studies due to its many above-mentioned capabilities [8–12]. Although these studies attempt to utilize VR systems to simulate the destabilizing environment with different kind of stimuli, most of the experiments did not focus on the natural ground walking locomotion in a virtual environment. Furthermore, we did not find any articles concerning the relationship between height exposure and gait stability. Most of the studies in this area have been focusing on the stationary postural sway as the dependent factor. For example, [8] compared the impact of real and virtual height on the postural stability of participants who were standing next to the edge of an elevated surface. However, they did not measure the impact of height on the gait stability of the subjects.

The proposed method of this study for measuring the gait pattern of the participants is to use a virtual environment and first-person character model enhanced with virtual legs to simulate a more realistic gait analysis on narrow, elevated surfaces. By collecting the time series data of the gait patterns from the participants' mounted trackers, this study sought to investigate potential differences between the average and variability of subjects' stride length and height due to elevation and increased realism.

## 2 Background

**2.1 Fall.** The leading cause of fatalities in the construction industry is fall, contributing to 38 percent of all fatal injuries happen every day at the construction sites[1]. Numerous research has been conducted recently to find the main reasons behind the disproportional number of fatalities associated with fall [2,13–18]. Slips, trips, and losses of balance are the most common factors contributing to falling [2,13,15]. Hsiao and Simeonov identified loss of balance as the leading cause of fall incidents. They stated that postural regulation of construction workers, especially those who work on elevated surfaces such as roofs or beams, is one of the leading factors in fall injuries and fatalities. Therefore, a closer look at the postural stability and influential factors affecting it can provide valuable insights into the fall-risk assessment.

**2.2 Postural stability.** Postural stability is defined as "the ability to maintain and control the body center of mass (CoM) within the base of support to prevent falls and complete desired movements [19]." Three central sensory cues for postural stability regulation are visual, vestibular, and somatosensory inputs. Compared to somatosensory and vestibular afferents triggered after the external exposure, visual input is considered a proactive mechanism of balance [2]. Paillard and Noe showed that expertise has a strong relationship with the visual input dependencies for postural regulation [20]. To accurately study the impact of visual input on the postural stability of the construction workers, we need to answer the following two questions: 1. How postural metrics can be measured, and 2. How do individuals respond to different visual perturbations affecting postural sway, especially those associated with height?

**2.3 Postural metrics.** Based on the definition of postural stability presented earlier, one way of measuring postural metrics is to measure the extent to which the center of the body deviates from the base (feet). Three principal reference planes are describing the anatomical motion of the body, transverse, frontal, and the median plane. The three sway directions namely frontier-posterior, mediolateral, and inferior sway happens respectively on the frontal, medial, and inferior planes. Biomechanical studies use maximum body sway (BS) in the frontier-posterior and mediolateral direction [21,22] and the root mean square of the total body velocity in frontier-posterior, mediolateral, and inferior direction [21] to measure the response of individual to different stimuli. The required data can be obtained by using the force plates or IMU sensors [23]. The other way of measuring the dynamic postural sway responses to external perturbations is to use Maximum Lyapunov exponent (Max LE) [24,25]. Lyapunov exponent directly measures the sensitivity of a dynamical system to minimal perturbations, considered to be a decent indicator of the level of chaos in the system [26]. By utilizing this method, the time series data captured by IMU sensors can then be investigated for any divergence reflecting the response of the dynamic system to the external perturbations. To efficiently find the reasons behind the postural instability of construction workers, identifying workers' responses to diverse types of external visual perturbations is of great importance. Accordingly, numerous studies have been conducted on human subjects to find out the impacts of different perturbations on the postural sway [3,19,27,28]. Interestingly, while postural stability is sensitive to the amplitude of these perturbations, the type of perturbation has far more impact on postural stability than the corresponding amplitude does [19]. Therefore, finding stimulus' characteristics is considered a critical step in fall-prevention techniques. Based on Hsiao and Simeonov critical review, the most significant environmental factors affecting postural stability are elevation, moving visual scenes, depth perception, visual ambiguity, and obstacle detection ability [2]. Since a high number of fatalities are due to falling from elevated surfaces, the current study will focus only on elevation as one of the prominent stimuli on postural stability and consequent loss of balance.

**2.4 Elevation, fear of height and VR.** The deficit of close visual contact due to the instigating sensory mismatch, and fear-related reactions, especially close to edges of elevated surfaces, are the two main characteristics of elevation leading to falling [2,29]. In many gaits and postural studies related to elevation and fear of height, pseudo-environment setup and stimuli simulations are the two common procedures to investigate the impact of the factors mentioned above. However, due to versatility of VR tools in generating virtual environment and various visual stimuli, and the unsafe nature of mechanical systems in provoking postural instability, VR systems have been widely used in fear-related behavioral and psychological responses [3,8,12,28,30]. Regenbrecht et al. scrutinized the concept of "presence" and explained that VR is a powerful tool in establishing an immersive environment to convey the feeling of presence to the user [30]. Wallach and Bar-Zvi showed that people suffering from flight phobia could be treated by using VR technology [12]. With an exposure-treatment approach, Rothbaum et al. measured the level of anxiety, avoidance, and attitudes of the participants before and after treatment with VR systems [28]. The outcome of the study showed a significant difference between the self-reported results of anxiety and avoidance of height of participants treated by computer-generated exposure, before and after treatments. In 2012, Celworth et al. compared the impact of height on the postural stability and anxiety level of subjects in a real and virtual environment. They used force plates to measure the center of body frequency and subsequently the stationary balance, and electrodermal activity (EDP) to measure anxiety. By sampling 17 young adults, while minimizing the inherent VR postural instability effect [7], they showed that virtual environment (VE) could simulate the height effect on the mediolateral and anterior-posterior balance of the subjects similar to the real height effect. They concluded that VR could be a useful study and rehabilitation tool for people with balance regulation deficits associated with fear or height. Numerous studies discussed the impact of gait speed and stride length on the overall gait stability and odds of falling in humans [31]. Step and stride length measures the distance between successive points of heel contact of opposite foot and same foot correspondingly. Based on the gait stability studies, the differences between the gait cycle patterns can be good indicators of the overall gait stability performance. During a simple walk the foot will experience periodic movement, which can be represented by the spatiotemporal acceleration and deceleration values. It is assumed that for a stable gait, the acceleration values should be following similar patterns and that the changes are periodically repeated. If there is any irregularity in this cyclic pattern, it indicates that there is an external (e.g., exposure to height) or internal disturbance (e.g., unstable CoM) in the system. Previous research suggests that people have a tendency to reduce their gait stride and speed when they suffer a preexisting fear of falling [32,33]. Interestingly, a higher level of walking disorder parameters--such as gait stride variability--can be affected by the fear of height as well [32].

In the light of foregoing research, the current study will focus on analyzing the time series data obtained in various VR/non-VR scenarios by retrieving the position of the HTC Vive trackers attached to the participants' feet. Furthermore, by quantitatively comparing these time-series data, this paper aims to present the ground walking experiments in VR as a promising tool for time-series data collection in different destabilizing environments, e.g., walking on an unguarded narrow elevated surface. To this end, the first-person character model developed in Unity will be enhanced with the virtual legs, helping the subject to walk on a virtual structural-beam path close to that he would perform in reality. Afterward, the role of VR legs and height in predicting the average value and the variability of gait stride parameters will be studied. By finding differences in gait pattern of the participants across these experiments, further data analysis for obtaining postural sways such as finding MaxLE or maximum BS becomes both possible and promising.

## 3 Methodology

To compare the impact of height on the gait stability of the 12 volunteer participants, two different experiments were designed (see Figure 1), each with two settings: one with no VR leg, and one with VR leg enhancement (see Table 1). In all the corresponding VR experiments, we attempted to minimize the inherent postural perturbations of VR headsets by simulating a VE as a static visual scene, removing any visual and physical distractions from the virtual environment, and minimizing the VR experiments durations (Horlings et al. 2009). To reduce the VR unfamiliarity, all the participants were instructed to stand, look, and walk for few steps in the VE for 1 minute total. The VR environment was the same environment that they would experience in the other parts of the data collection. After the completion of this part, the participants were equipped with three HTC Vive trackers and an HTC Vive headset. In the first trial, the 'immersed' beam path was presented to the participants while they were able to see their virtual legs. The immersed path consists of a closed triangular polygon constructed by three connected structural beams, with no elevation (figure 1-a). The participants were instructed to walk on the path at a comfortable speed. During this trial and others that followed, the coordinates of the trackers attached to the participants' feet were retrieved with the refresh rate of 60 times per second. The procedure for the second trial was identical to the first trial except that the VR leg enhancement was not available to the participants. In the third trial, the participants were subjected to height by locating the triangular beam path on the 17th floor of an unfinished building (figure 1-b). The virtual environment remained unchanged save for the height. Similar to the first trial, the participants were able to see their VR legs. In the last trial, the participants were asked to walk again on the same elevated path again. However, the third trial excluded virtual legs from the VR model. In other words, the participants were not able to see their virtual legs as they had been in the first and second trials.

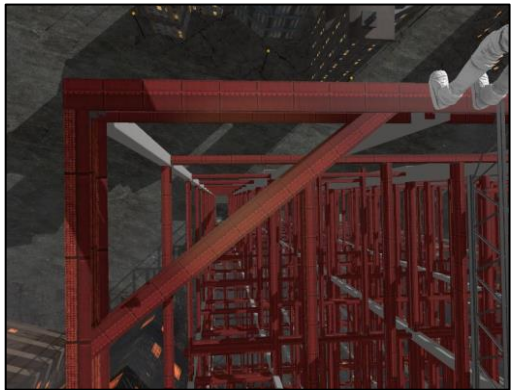

(a)

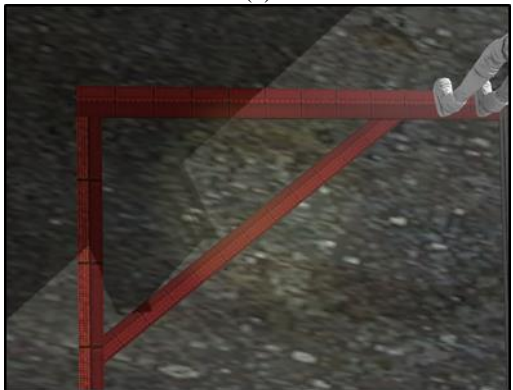

(b)

Figure 1 Two different scenarios namely, real office at the top left, the same virtual office at the bottom left, the virtual elevated path at the top right, and virtual narrow elevated path at the bottom right

Table 1 Experiment configuration

| Setting | Scenarios | |
|---|---|---|
| No virtual legs | VR path on the ground | Elevated VR path |
| Virtual legs | | |

## 4 Results and Analysis

Gait analysis often includes the spatial and temporal measurement of a gait cycle so that the interesting factors can be inferred by comparing the differences between the measurements. Therefore, the focus of this study was on the gait parameters; namely, average speed, length of step and stride, and their variabilities. As mentioned previously, the x, y, and z coordinates of each tracker (attached to participants' feet) were recorded during each trial. Figure 2 shows a sample gait pattern plotted by using one of the participant's right tracker position dataset. The solid lines showed trials with models enhanced with virtual legs and the dashed lines showed those trials without any enhanced virtual legs. Similarly, those lines accompanied by the astride sign '*' represent trials conducted in the presence of virtual height, while those with no '*' represents trials without the presence of elevation. To study the impact of our two purposed factors—namely, height and virtual legs--we conducted series of two-tailed paired T-tests between relevant groups divided based on the presence of the focused factor. The results of the paired T-test are valid as long as the distribution of the data for each group is normal, and each participant is compared with himself/herself. To test the normality of the data, the Anderson-Darling test was performed for each comparison executed between the two paired groups (Anderson and Darling 1952). For each comparison case, the null hypothesis was not rejected. More specifically, the results of the 'normality tests' showed that the distribution of the paired groups data is normally distributed.

By statistically comparing the gait durations across all scenarios, we found out that (1) LEGS: by enhancing the participants' virtual models with virtual legs, the duration of the ground walking trials increased significantly compared to cases in which the virtual legs were not present (P value= 0.0004). Also, the paired T-test showed that the presence of VR legs noticeably decreased the average gait stride height on the non-elevated walking path. Interestingly, no significant increase in the duration of the trials performed at height was observed while the effect of virtual legs was considered, and (2) HEIGHT: while the participants were able to use their virtual legs in the manifestation of height, the average stride length of the participants decreased significantly (P value= 0.00004). Even without the virtual legs, a similar height effect was detected (P value=0.015). In addition, the duration of the trials performed with (P value=0.03) and without (P value=0.005) virtual legs was noticeably increased once the subjects were exposed to height. No prominent difference in the stride heights was detected with or without the presence of height. However, the stride height variability was meaningfully increased when the participants, equipped with the enhanced model, were subjected to height (P value=0.02).

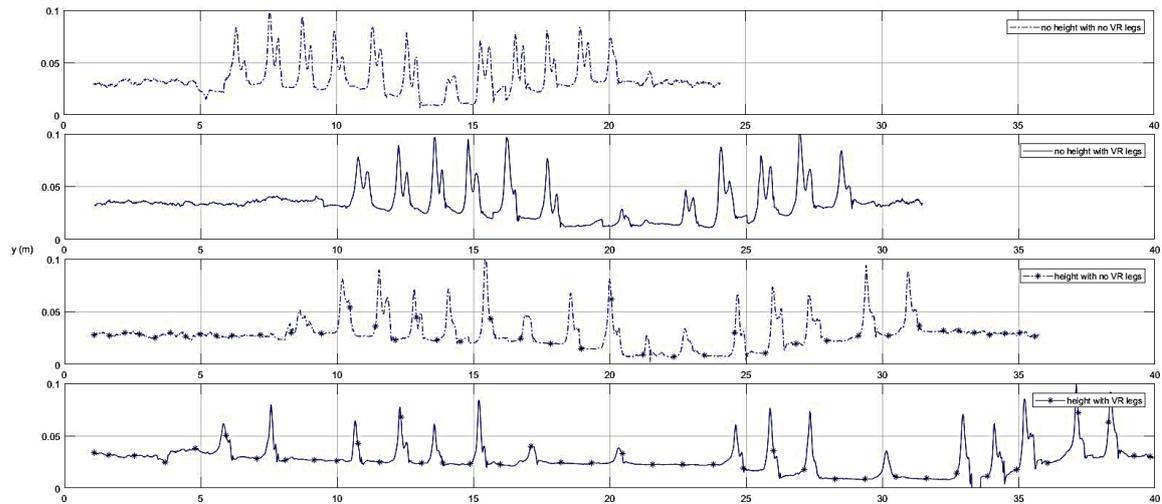

Figure 2 Comparison of the gait patterns of a sample participant during the four trials. The continues lines and those accompanied by the Astrid sign '*' represent the usage of VR legs and the presence of VR height respectively.

Table 2 Two-tail paired T test's significant level of mean differences between each pair of groups (P values)

|  |  | average |  | variability (sd) |  |
| --- | --- | --- | --- | --- | --- |
|  | Exp. Duration | stride length | stride height | stride length | stride height |
| no L[1] (H vs. no H) | 0.0052* | 0.0145 | 0.7609 | 0.6631 | 0.0678 |
| L (H vs. no H) | 0.0305* | 0.0000** | 0.6554 | 0.4850 | 0.0281* |
| no H(L vs. no L) | 0.0004** | 0.4464 | 0.0227* | 0.2518 | 0.7381 |
| H(L vs. no L) | 0.2912 | 0.2968 | 0.2634 | 0.0884 | 0.7290 |

[1] L: VR legs, H: height
* Sig. at 0.05 level
** Sig. at 0.01 level

## 5 Discussion and Conclusion

By analyzing the data regarding the basic spatiotemporal variables of stride length and speed, we detected a noticeable stride length difference across different scenarios. More precisely, when we focused on height as the influential factor, we observed that the number of strides significantly increased while the average stride length significantly decreased. Hence, these reductions of participants' gait speed and stride length imply that the participants attempted to walk more "carefully." The decrease in gait speed and stride length between ground and elevated walking is in line with Sheik_Nainar and Kaber's findings [34]. Also, Schniepp et al.'s experiment results [35] demonstrated the same bodily reactions from participants subject to destabilizing environments. The decrease in stride length as a result of participants being asked to walk on the elevated path confirms Schniepp et al.'s previous results, with the only difference being a real elevated path for their experiments as compared to a VR simulation in ours.

Another interesting result was the reduction in the average stride length combined with an increase in both the average stride height and the duration of the trials once the participants had access to the enhanced VR model for the ground walking trials. In other words, with the presence of virtual legs, the participants walked on the path instead of just 'following the path.' This would imply that the use of responsive and real-time tracking virtual legs will increase the level of realism for the participants. However, a similar statistically significant result was not spotted for the elevated walking trials. The average participant's stride length did not decrease significantly (P-value=0.296). This finding perhaps indicates that height was the dominant factor here, reducing the average stride length even without the presence of the virtual legs.

Based on the result of table 2, the stride height variability was significantly affected by height as well (P-value=0.0281). The standard deviation of the stride height was lower for the data obtained from the ground walking trial with the presence of virtual legs compared to those extracted from the elevated path that also included the presence of virtual legs. This higher level of walking disorder parameters can be affected by the fear of height [32]. The thought-provoking part of these findings was that, in the absence of virtual legs, there were no significant variability differences between the two datasets pertaining to ground walking and walking on the height. Whether the cause of this phenomenon was the participants' attempts to place their feet exactly on the beam or the participants' experience of changes in their visual and vestibular systems, we cannot know for sure. However, it is quite apparent that the presence of virtual legs has an impact on the gait parameters, even for the higher level of walking disorder parameters. Therefore, we conclude that walking on the virtual structural beams without being able to look at one's virtual legs is less realistic than when someone walks with virtual legs in view.

Overall, these findings show that VR, along with the advanced virtual models, was successful in inducing the height effect on the basic spatiotemporal variables related to gait. Since these variables are essential to postural sway and gait stability measurements, the finding of the current study could be a good start in designing efficient, height-related experiments for construction workers' posture and gait assessments with fewer safety concerns. In addition, many of these virtual experiments can be integrated with other evolving construction safety studies such as eye-tracking and risk perception [36,37], cultural factors and risk perception [38], and intoxication at construction sites [39].